\documentclass[a4paper,12pt]{article}
\usepackage{geometry}
\usepackage{amsmath,amssymb}
\usepackage{graphicx}
\usepackage{caption}
\usepackage{subcaption}
\usepackage{float}
\usepackage{array}
\usepackage{authblk}
\usepackage{booktabs}
\usepackage{setspace}
\graphicspath{{figures/}{images/}}

\usepackage[numbers,sort&compress]{natbib}
\usepackage{hyperref}
\hypersetup{
	colorlinks=true,
	citecolor=blue,
	linkcolor=black,
	urlcolor=blue
}

\renewenvironment{abstract}
{\quotation\noindent\textbf{Abstract:}}
{\endquotation}

\title{Simple Survival End Point Calculator}
\author{Shaodong Wenren, Dingyan Cao, Zhiheng Jiang, Chenyu Li}
\affil{East China Normal University, Shanghai, 200062}
\date{}

\begin{document}
	
	\maketitle
	
	\begin{abstract}
		\noindent
		In the study of survival endpoints, the size of the sample is crucial. The calculation of the sample size for survival endpoints was first proposed and presented by Freedman based on the logrank statistic. Later, many statisticians gradually introduced multiple factors such as dropout rate and delayed onset, proposed various estimation methods including the lrstat method, and developed corresponding statistical programs. However, the application of professional statistical functions is relatively complex. Based on this, this paper has developed a corresponding visual web application on the basis of the R package lrstat and powerSurvEpi, and tested its functions and program accordingly. The test results are basically consistent with the case, proving the effectiveness of the application.
		
		\noindent \textbf{key words}:logrank test, lrstat test, survival endpoint, sample size, R language
	\end{abstract}
	
	\section{Introduction}
	
	The endpoint of survival, in the study of biostatistics, refers to the occurrence of a certain event. For example, in the study of survival time for advanced cancer patients, death is often used as a survival endpoint. However, an obvious issue is that not all survival endpoints can be accurately measured. Taking cancer patients as an example, death events may still not occur at the end of the experiment. This phenomenon is called censoring. Due to the influence of censoring and long cycles, the calculation of sample size for survival endpoints must take into account the time dimension and sample censoring, which makes the calculation of sample size for survival endpoint studies particularly complex.
	
	Moreover, due to the longer experimental period and the requirement for full process tracking, the calculation of sample size for survival endpoints is also influenced by many other factors. The experimental samples may be enrolled in batches over several months, and the enrolled samples may suddenly lose contact after a certain point in time. The combination of these factors makes determining the sample size required for experiments with survival endpoints a relatively complex and difficult task. Based on this, we developed a simple R software for quickly calculating the sample size required for survival endpoint experiments according to given conditions.
	
	\section{Relative Models}
	
	In order to specifically solve the above problems and implement the model into a simple app, after consulting relevant literature, the author chose the Lakatos method based on logrank statistics and its extensions, including lrstat, as the theoretical basis.
	
	\subsection{logrank method}
	
	The logrank method was proposed by L.S. Freedman in 1982 \cite{Freedman1982}. At that time, several authors pointed out after detailed analysis that there was a problem of insufficient patient numbers in a considerable number of studies in the field of cancer, which led to distorted and even erroneous conclusions. In order to fully solve the problem of minimum sample size, Freedman proposed the logrank test method and created corresponding tables.
	
	Suppose that two treatments give rise to survival rates of $P_{1}$ and $P_{2}$ , and hazard ratio $\theta$, the quantities $P_{1}$, $P_{2}$, and $\theta$ are related by: 
	\begin{align}
		\theta = \frac{lnP_{1}}{lnP_{2}}
	\end{align}
	
	The total number of events $d$ in both series, needed to be observed in a trial is:
	\begin{align}
		d = (z_{1}+z_{2})^{2}(\frac{1+\theta}{1-\theta})^{2}
	\end{align}
	
	where $z_{1}$ is the normal deviate corresponding to the particular significance level employed in the logrank test and $z_{2}$ is the normal deviate corresponding to the required power.
	
	By arguing conditionally on the set of patients at risk before each event and letting $\phi_{i}$ denote the ratio of patients at risk in the two groups before event i(i = 1,. . . , d), then we get  the expectation and variance: 
	
	\begin{align}
		E = \frac{\sum_{i=1}^{d}[\frac{\phi_{i}\theta}{1+\phi_{i}\theta}-\frac{\phi_{i}}{1+\phi_{i}}]}{\sqrt{\sum_{i=1}^{d}\frac{\phi_{i}}{(1+\phi_{i})^{2}}}} \\
		V = \sum_{i=1}^{d}[\frac{\frac{\phi_{i}\theta}{(1+\phi_{i}\theta)^{2}}}{\sum_{i=1}^{d}[\frac{\phi_{i}}{(1+\phi_{i})^{2}}]}]
	\end{align}
	
	Finally, we get the formula for sample size: 
	\begin{align}
		n = \frac{d(1+\phi)}{\phi(1-P_{1})+(1-P_{2})}
	\end{align}
	
	This is the basic logrank model.
	
	\subsection{Lakatos method and its expansion lrstat}
	
	The logrank test has been widely applied since its proposal, especially achieving great success in clinical trials. However, with the increase of application fields, the logrank test has also been tested by complex and ever-changing real-world environments. For example, when the hazard ratio is between 0.5 and 2, the logrank method is basically completely effective. However, once the hazard ratio exceeds this value, the effectiveness of the logrank method will decrease; In addition, Freedman previously proposed a flawed hypothesis that the analysis was conducted after a fixed period of time after the last sample entered the study, and the follow-up data collected after analyzing this time point were discarded. This leads to an overestimation of the sample size and can cause serious bias in some cases. Based on this, researchers Lu et al. further improved and extended the Lakatos method, alleviating the above problems and providing appropriate answers to general follow-up plans, non inferiority tests, and predicting the number of events within calendar time. These improvements ultimately formed the extended Lakatos method, lrstat\cite{lrstat2016}.
	
	\section{Critical function and calculation method}
	
	The software developed by the author in R language is a code implementation of the logrank test proposed by Freedman and the extended Lakatos method. This software supports two methods for calculation, each with corresponding parameters to choose from.
	
	\subsection{Freedman method}
	
	The Freedman method, also known as the classic logrank test, is applicable to classical proportional hazards models. This model can select four main parameters, from top to bottom: firstly, the calculation direction, which can choose to automatically calculate the sample size based on the given power, or choose to calculate the power under this condition based on the given sample size; Next is the setting of statistical parameters, including significance level $\alpha $, efficacy $1-\beta $, one-sided or two-sided test, and the proportion of samples randomly assigned to the experimental group and the control group, which is $n_{E}/n_{C}$; Next is to set the time parameters for the study, including enrollment time, total time required for all samples to enter the study process, study time, and duration of the study; The final step is to set survival and effect parameters, including the survival time of the control group and the risk ratio between the experimental group and the control group.
	
	\subsection{lrstat method}
	
	The lrstat method is an extension and improvement of the Freedman logrank test. Its parameter settings are basically the same as the Freedman method, but there are differences in some parameter settings. Firstly, the statistical parameter settings are different from the Freedman method. The lrstat method supports non inferiority tests, which are tests with a hazard factor greater than 1. Therefore, when setting parameters, you can choose the non inferiority test and select the specific size of the non inferiority coefficient in the box below; In addition, the lrstat method also supports more advanced settings, including the following: delayed onset time, setting a delayed onset time during which the two groups have the same risk, and then the experimental group's risk will decrease; Enrollment rate, monthly sample size entering the study, annual dropout rate, annual sample size terminating follow-up, and Fleming Harrington test function are used to differentiate between the early, middle, and late stages.
	
	\subsection{Results visualization}
	
	The software displays results on four pages: sample size calculation, survival curve, event prediction, and method comparison.
	
	The sample size calculation page displays all parameters and related calculation results mentioned in sections 3.1 and 3.2, including experimental group sample size, control group sample size, expected number of events, control group event probability, and experimental group event probability; The survival curve page will provide survival curve graphs for the experimental group and the control group, with the x-axis representing study time and the y-axis representing survival probability; Draw the cumulative event prediction for each calendar time point on the event prediction page (based on the lrstat parameter), displaying the cumulative number of participants and the expected number of events; The method comparison page displays a comparison of the calculation results of two methods under the premise of a unified time scale, including total sample size N, number of events D, study duration, non proportional risk, delayed effect DTE, and lost to follow-up treatment.
	
	\subsection{Calculating logic}
	
	This web app uses lrstat and powerSurvEpi for calculations. When using the LR method, the main function used is lrsamplesize, which calculates the sample size based on the test performance, lrpower, Calculate the testing efficiency based on the sample size; When using the Freedman logrank method, the main functions used are ssizeCT.dault and powerCT.dault, which are used to calculate sample size and test performance, respectively\cite{lrstat}\cite{powerSurvEpi}.
	
	\section{Application test and example analysis}
	
	The author specifically selected three cases for case analysis, two simulated cases for testing functionality, and one actual case for calculation. The test results are following. 
	
	\subsection{lrstat simulation}
	
	This is a simulation case used to test the operation of software functions.
	
	\begin{table}[H]
		\centering
		\caption{lrstat parameters}
		\begin{tabular}{ll}
			\toprule
			\textbf{Method} & lrstat (Lu 2021) \\
			\textbf{Result} & sample size \\
			\textbf{Sample size N} & 285 \\
			\textbf{Incidents expectations D} & 247 \\
			\textbf{Research time(months)} & 65.8 \\
			\textbf{Power} & 80\% \\
			\textbf{Control group median survival time(month)} & 12 \\
			\textbf{Target HR} & 0.7 \\
			\textbf{Non-inferiority margin HR0} & 1 \\
			\textbf{Duration of enrollment(month)} & 18 \\
			\textbf{Follow-up duration(month)} & 48 \\
			\textbf{Enrollment rate(person/month)} & 10 \\
			\textbf{Delayed onset(month)} & 0 \\
			\textbf{Annual loss rate} & 0.05 \\
			\textbf{FH rho1/rho2} & 0/0 \\
			\textbf{Randomization ratio E:C} & 1:1 \\
			\textbf{Alpha} & 0.05 \\
			\textbf{Type of test} & two-sided \\
			\bottomrule
		\end{tabular}
	\end{table}
	
	\begin{figure}[H]
		\centering
		\includegraphics[width=0.9\textwidth]{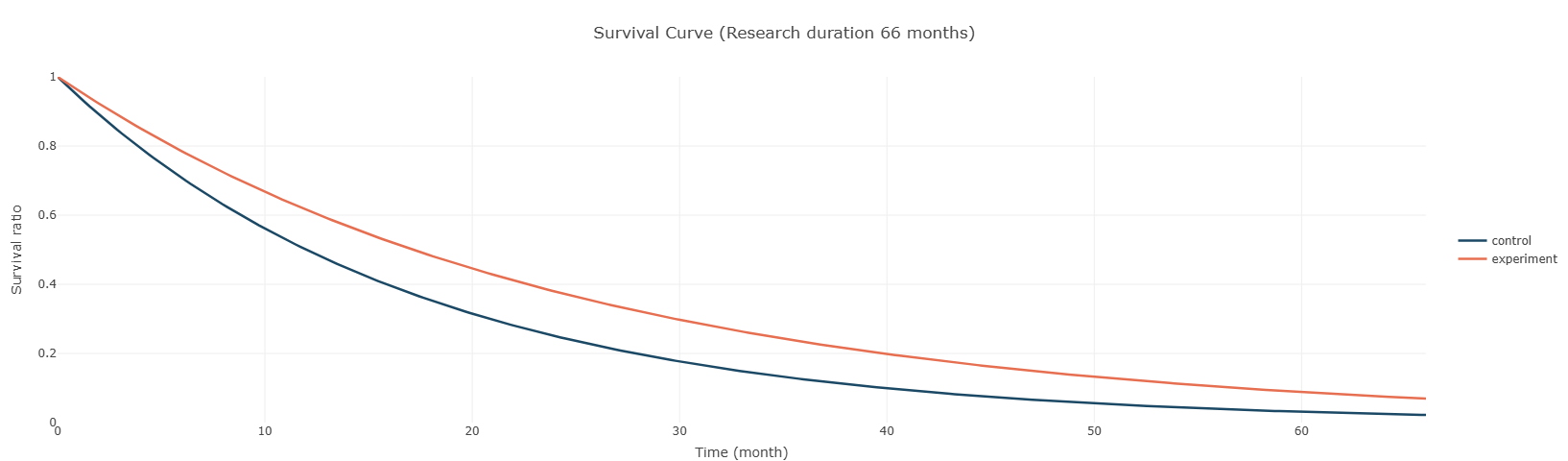}
		\caption{4.1 Survival Curve}
	\end{figure}
	
	\begin{figure}[H]
		\centering
		\includegraphics[width=0.9\textwidth]{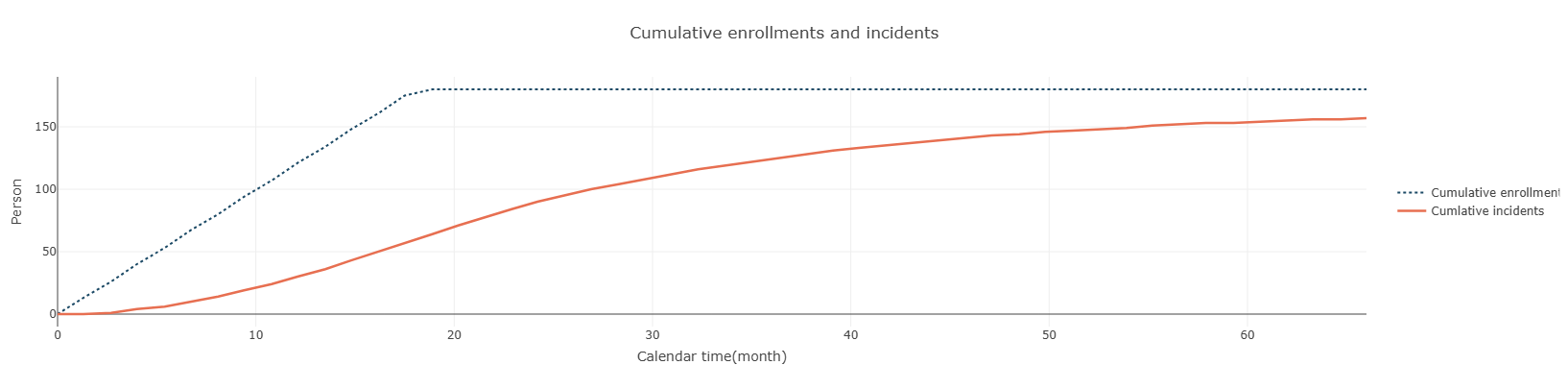}
		\caption{4.1 Cumulative Enrollments and Incidents}
	\end{figure}
	
	\subsection{powerSurvEpi simulation}
	
	\begin{table}[H]
		\centering
		\caption{powerSurvEpi parameters}
		\begin{tabular}{ll}
			\toprule
			\textbf{Method} & powerSurvEpi (Freedman 1982) \\
			\textbf{Result} & sample size \\
			\textbf{Sample size N} & 696 \\
			\textbf{Experiment sample size} & 348 \\
			\textbf{Control sample size} & 348 \\
			\textbf{Incidents expectations D} & 637 \\
			\textbf{Power} & 80\% \\
			\textbf{Total research time(month)} & 48 \\
			\textbf{Control group median survival time(month)} & 12 \\
			\textbf{Target HR} & 0.8 \\
			\textbf{Control incidents probability} & 0.9375 \\
			\textbf{Experiment incidents probability} & 0.8912 \\
			\textbf{Alpha} & 0.05 \\
			\textbf{Type of test} & two-sided \\
			\bottomrule
		\end{tabular}
	\end{table}
	
	\begin{figure}[H]
		\centering
		\includegraphics[width=0.9\textwidth]{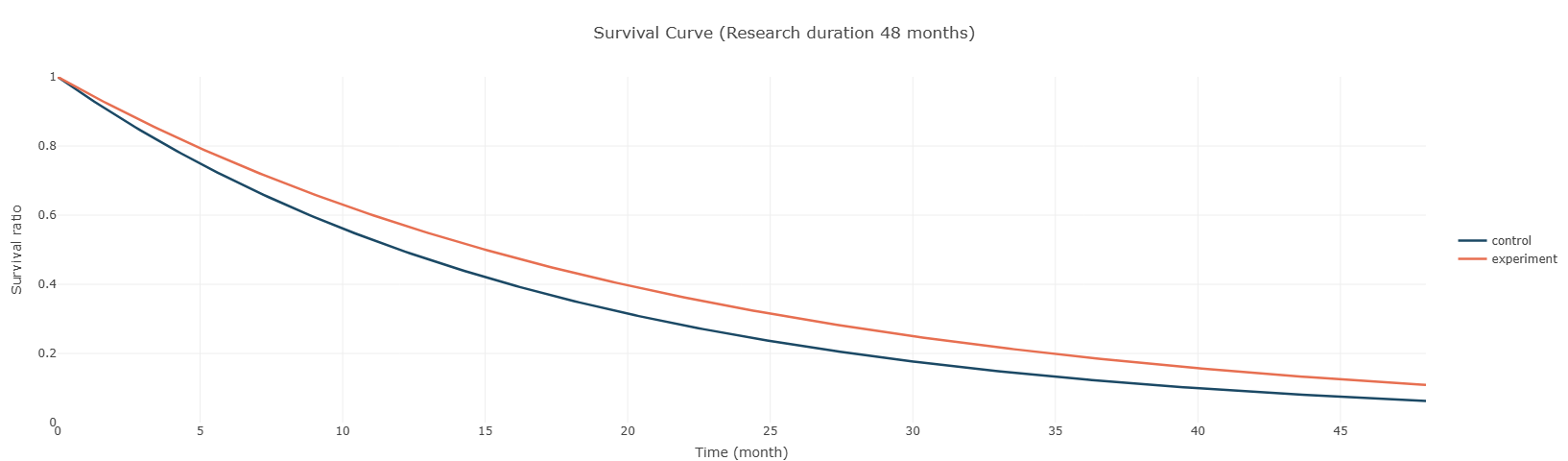}
		\caption{4.2 Survival Curve}
	\end{figure}
	
	\begin{figure}[H]
		\centering
		\includegraphics[width=0.9\textwidth]{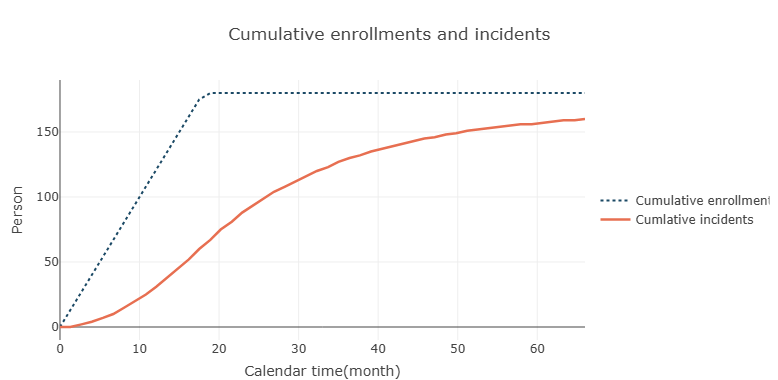}
		\caption{4.2 Cumulative Enrollments and Incidents}
	\end{figure}
	
	\subsection{Retinitis Pigmentosa}
	
	\begin{table}[H]
		\centering
		\caption{Retinitis Pigmentosa parameters}
		\begin{tabular}{ll}
			\toprule
			\textbf{Method} & lrstat (Lu 2021) \\
			\textbf{Result} & sample size \\
			\textbf{Sample size N} & 320 \\
			\textbf{Incidents expectations D} & 197 \\
			\textbf{Research time(months)} & 83.9 \\
			\textbf{Power} & 80\% \\
			\textbf{Control group median survival time(month)} & 36 \\
			\textbf{Target HR} & 0.67 \\
			\textbf{Non-inferiority margin HR0} & 1 \\
			\textbf{Duration of enrollment(month)} & 36 \\
			\textbf{Follow-up duration(month)} & 48 \\
			\textbf{Enrollment rate(person/month)} & 5.55 \\
			\textbf{Delayed onset(month)} & 0 \\
			\textbf{Annual loss rate} & 0.02 \\
			\textbf{FH rho1/rho2} & 0/0 \\
			\textbf{Randomization ratio E:C} & 1:1 \\
			\textbf{Alpha} & 0.05 \\
			\textbf{Type of test} & two-sided \\
			\bottomrule
		\end{tabular}
	\end{table}
	
	\begin{figure}[H]
		\centering
		\includegraphics[width=0.9\textwidth]{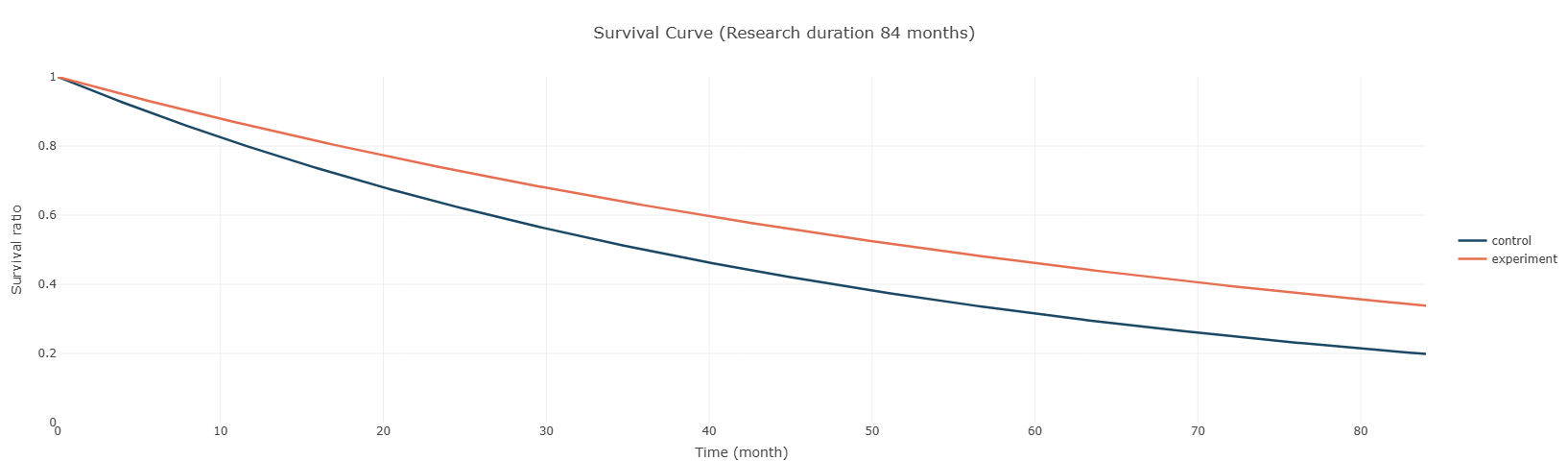}
		\caption{4.3 Survival Curve}
	\end{figure}
	
	\begin{figure}[H]
		\centering
		\includegraphics[width=0.9\textwidth]{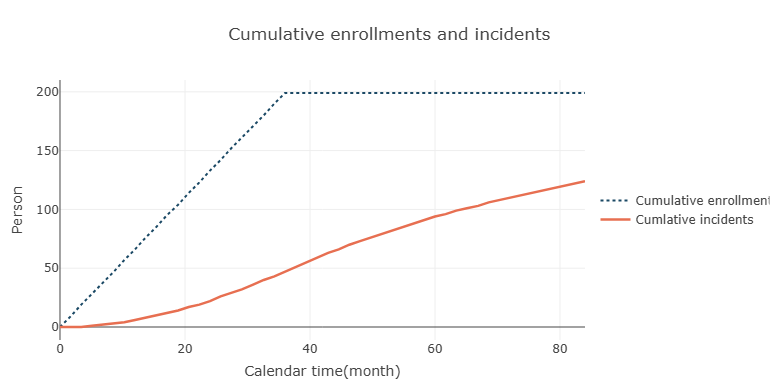}
		\caption{4.3 Cumulative Enrollments and Incidents}
	\end{figure}
	
	\section{Conclusions and Discussions}
	
	After our testing, the software supports both the standard Freedman method and the improved lrstat method. It can select the corresponding significance level, power, and other related parameters according to its own needs. It can adjust the relevant parameters of the experimental group, control group, and experimental population according to different experimental conditions, and provide different options for different requirements. In the display of calculation results, there are not only quantitative results, but also visual charts, including survival curve graphs, event prediction graphs, and method comparison graphs. Thus, the software has basically achieved the functions we requested.
	
	However, the app still has some shortcomings, mainly including the following two points: firstly, it only supports two methods, which is quite limited for non-standard applications; Secondly, the drawn graphics lack a language function. These issues will become the direction for further improvement, and our upcoming updates will revolve around these issues.
	\section*{Availability}
	The application is available as the R package \textit{survSampleSize} on CRAN (https://CRAN.R-project.org/package=survSampleSize) and on GitHub (https://github.com/wettlinmalfa629-hue/survSampleSize).
	\bibliography{references}

\end{document}